# Enhanced sensitivity distributed sensing of magnetic fields in optical fiber using random Bragg grating

Antoine Leymonerie,[1] Jean-Sébastien Boisvert,[2] Léonie Juszczak,[1] Sébastien loranger[1,*]

[1]Electrical Engineering, Polytechnique Montreal, 2500 ch de Polytechnique, Montreal H3T 1J4, Canada
[2]Engineering Physics, Polytechnique Montreal, 2500 ch de Polytechnique, Montreal H3T 1J4, Canada
*sebastien.loranger@polymtl.ca

We show that the use of Random Optical Grating using UV Exposure (ROGUE) can significantly reduce the noise floor of an optical frequency domain reflectometry (OFDR) measurement of Faraday rotation in the polarization. We compare it with unexposed spun fiber which shows a S/P minimum ratio (signal noise floor) 20 dB higher than when using our ROGUE. High sensitivity magnetic field measurements are achieved by spatially filtering (setting a gage length) the derivative of the S/P ratio's evolution. Example of a calibrated electromagnet spatially resolved B-field measurement is demonstrated, which can measure fields down to 10 mT with 10 cm spatial resolution. The potential for current sensing using such ROGUE apparatus is discussed and simulation shows a noise floor of ~1 A with 40 probing loops spatial resolution.

Optical fiber sensors have shown great advantage in many fields thanks to their remote sensing capability, their immunity to electromagnetic noise, their tolerance to harsh environmental conditions and their high sensitivity [1]. In addition, optical fiber can be used for distributed sensing, offering spatially resolved continuous sensing data. This can be extremely useful for high-precision modeling, such as in shape sensing [2], and to view localized events, such as faults, which is crucial in many safety monitoring applications [3]. In this paper, we are interested in distributed magnetic field measurements, which can be extended to high-current monitoring (also known as all-fiber optic current transformer (AFOCT)) [4]. Distributed magnetic field is critical in applications where the shape of the field is important to know and model, such as in Tokamaks [5], industrial magnets in motors [6] and medical MRI systems [7]. Distributed current monitoring can be useful in high-current industrial applications to monitor several loads in a single measurement or to detect faults before catastrophic failures.

Highly sensitive optical fiber magnetic sensors [8] have been demonstrated through strain measurement induced by magnetostriction [9, 10] or through phase measurement of a surrounding fluid [11]. In both cases, the fiber needs to interact with a magnetic responsive solid or fluid, requiring complex sensor fabrication procedure that can hardly be scaled to long lengths. Another means of measurement is through Faraday rotation [12], where a current is measured by integrating the polarization rotation after a significant number of fiber loops around the current source. Although such sensors can be very sensitive, they are limited to a single integrated measurement with no spatial information. Palmieri et al. have shown distributed current sensing by polarization analysis through optical frequency domain reflectometry (OFDR) of Rayleigh back-scattering in standard fiber [13], which resulted in a noise floor of ~100 A for 11 m of optical fiber length (40 loops) resolution.

OFDR distributed sensing of Rayleigh back-scattering has shown many sensing applications (temperature, strain, vibration & shape), but always suffers from very weak signal coming back to the interrogator. Various techniques have been proposed to increase such signals, such as UV exposure of fiber [14], continuous weak fiber Bragg grating (FBG) writing [15] or FBG arrays [16]. This latter technique was demonstrated as distributed magnetic sensor in Tokamaks [5]. However, continuous FBGs have the disadvantage of limited length due to having all the signal being reflected, while FBG arrays suffer from large manufacturing cost and blind spots between gratings. A recently proposed solution is to use incoherent FBG, a Random Optical Grating using UV or fs exposure (ROGUE), which is not only very inexpensive to manufacture (being noise-generated), but also offers a large-bandwidth significantly enhanced return signal while maintaining a long operation length [17]. In this paper, we demonstrate how a ROGUE's enhance signal can significantly lower the noise floor of an OFDR magnetic measurement, thus making possible the measurement of low field with cm-scale spatial resolutions.

ROGUEs are written using a Talbot interferometer, as shown in Fig. 1a. In this setup, the fiber is moved continuously under a fringe pattern. The intrinsic noise of the motor induces FBG writing over a coherence length equal to the spot size. Hence, the natural bandwidth of the ROGUE is inversely proportional to the spot size. The strength and bandwidth of the ROGUE can be tuned through coherence control by sending a random noise amplitude to a piezoelectric element moving the phase-mask [17]. Fig. 1b shows the total spectra of the 40 cm ROGUE used in this paper.

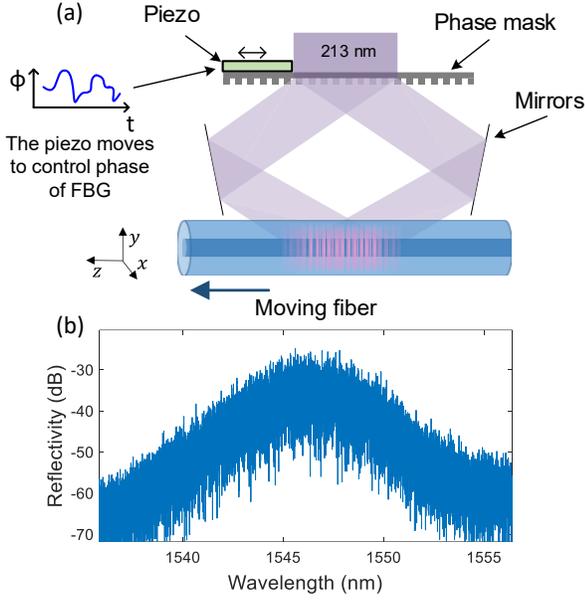

Fig. 1. (a) ROGUE fabrication setup using a Talbot interferometer to set fringe period and a piezo to control phase. Coherence is controlled by the ratio of sawtooth wave (to move the fringe pattern synchronously with fiber) and a noise function driving the piezo. (b) Resulting broadband ROGUE total reflectivity spectra used in this paper.

The principle of operation relies on the ability of an OFDR optical back-reflectometer (OBR) to measure the spatial distribution of 2 orthogonal polarization states: $S(z)$ and $P(z)$ (linearly polarized at the detectors). If the state entering the fiber under test (FUT) at $z = 0$ is linearly polarized, then the polarization will rotate by $\theta = VBz$ as it propagates in the fiber under a magnetic field $B$ along its z-axis, where $V$ is the Verdet constant of the material ($V \approx 0.6$ rad/(T·m) for silica optical fiber [13]). This optical activity is caused by magnetically induced weak circular birefringence. Therefore, it is crucial for the linear birefringence of the medium to be either null or very small. This is why magnetic or current sensing is typically done in spun fiber, which offers circular birefringence. Through the Faraday rotation principle, the reflected light will continue to rotate in the same direction, yielding a total rotation angle of $\theta = 2VBz$. If we extinguish one polarization to its minimum at $z = 0$ (S-polarization in this paper) by setting the input state with a polarization controller (PC), then the distributed signal will be:

$$\frac{S(z)}{P(z)} = \tan\theta = \tan\left(2V\int_0^z B(z')dz\right). \quad (1)$$

The field can then be expressed as the derivative of the amplitude signal of the S/P ratio:

$$B(z) = \mu H(z) = \frac{d}{dz}\left(\frac{1}{2V}\tan^{-1}\left(\frac{S}{P}\right)\right), \quad (2)$$

where $H$ is the field strength (in A/M) and $\mu$ is magnetic permeability ($\mu_0$ for silica). The effect of weak linear birefringence will add a predictable error along the length which can be corrected, as long as $z < L_{bi-lin}/4$, where $L_{bi-lin}$ is the linear birefringence beat-length, beyond which the optical activity will be periodically reversed. In this paper, the lengths analyzed are much smaller than $L_{bi-lin}$ and can be ignored. It should be noted that a highly-resolved OFDR signal is generally highly noisy with a single-point resolution of $\delta z \sim 100$ μm. Reducing the noise floor to acceptable level is done through post-analysis spatial filtering and large-area derivation. The nature of this filtering is subject to optimization but will always yield a certain spatial resolution $\Delta z$ (considered as the full-width at half-maximum (FWHM) of a simulated point-spread function (PSF) of a highly localized magnetic field). A simple spatial filtering would consist of a large-area derivative, discretized as:

$$b_i = \frac{\tan^{-1}\left(\frac{S_{i+M}}{p_{i+M}}\right) - \tan^{-1}\left(\frac{S_{i-M}}{p_{i-M}}\right)}{4VM\delta z}, \quad (3)$$

where $b_n$ is the discretized magnetic field from Eq. 2, $M$ is the number of points used in the averaging and $\delta z$ is the spatial spacing between points. In this case, $\Delta z = M\delta z$. For this simple case the standard deviation ($\sigma_B$) noise floor would be:

$$\sigma_B = \frac{\sigma_{OBR}}{V\Delta z R} \cdot \begin{cases} 1 & S \ll P \text{ or } P \ll S \\ \sqrt{2} & S \sim P \end{cases}, \quad (4)$$

where $\sigma_{OBR}$ is the OBR signal noise floor (red dashed line in Fig. 2b) for either P or S polarization (we suppose the same noise floor for both detectors) and $R$ is the total signal ($R = S + P$). We can immediately see the advantage of ROGUEs, as the relative noise $R/\sigma_{OBR}$ is much stronger (>40 dB gain) compared to non-exposed spun fiber or standard fiber (Fig. 2b). In the simple single-filtering model of Eq. 3, the noise floor of the magnetic field is inversely proportional to the chosen resolution.

This can be further improved by using higher-order filtering technique. In this paper, we use a 3-step filtering: a logarithmic pre-smoothing (moving average), a large area derivative and a post-smoothing. These steps can be represented by the following discretization:

$$r_k = \tan^{-1}\left(\exp\left(\frac{1}{2M}\sum_{i=k-M}^{k+M} \ln\left(\frac{S_i}{p_i}\right)\right)\right), \quad (5a)$$

$$g_k = \frac{r_{k+M} - r_{k-M}}{2M\delta z}, \quad (5b)$$

$$b_n = \frac{1}{4VM}\sum_{k=n-M}^{n+M} g_k. \quad (5c)$$

From a numerical analysis of the noise, this 3-step filtering gives a $\Delta z^{-1.7}$ dependance for the resulting magnetic field noise and decreases it by a factor of 65 for a 4 cm resolution compared to Eq. 3.

To demonstrate distributed magnetic measurement using ROGUE, the setup shown in Fig. 2a is used. The interrogation system is composed of a commercial OFDR system from LUNA (OBR4600) connected to a polarization controller (PC) to adjust the polarization at the input of the FUT, which is optimized to have the highest extinction of the S-polarization at zero-field. A higher extinction ratio can be reached with the ROGUE (26 dB vs 11 dB for unexposed fiber) as the suppressed polarization's amplitude is much farther from the OBR noise level. The residual fixed polarization contamination signal above noise is not problematic, as we calculate the derivative. A PM delivery fiber is used between the interrogation system and the measurement location (3-meters away) to ensure bending insensitivity of the delivery fiber. The FUT is composed of a section of spun fiber (section A) and a 40 cm ROGUE (section B) written in hydrogen loaded SMF-28. The FUT is then placed in an electromagnet in axial configuration. Two calibrated electromagnets (using Hall-efect sensors) were used for tests: a double water-cooled copper-coiled magnet in Helmholtz configuration which can reach 400 mT at its center (Mag1) and a superconducting magnet capable of reaching 5 T (Mag2).

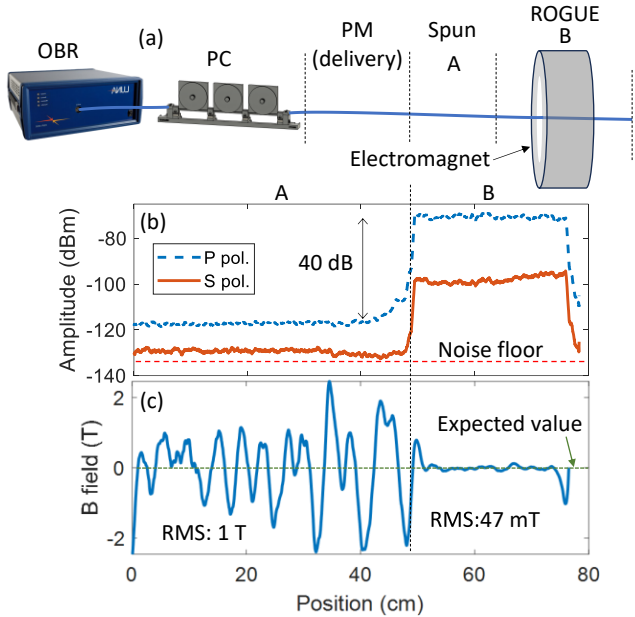

Fig. 2: (a) Measurement setup. PM fiber is used for stable delivery to the test regions. (b) OFDR spatial measurement of the P and S polarization amplitude signals (spatially averaged for visualization). The average noise floor of the OBR (as measured after end-of-fiber) is shown with the red dashed line (c) Comparative analysis of the noise between a non-exposed spun fiber and a ROGUE after applying Eq. 5 for a spatial filtering of 2 cm (FWHM) with no magnetic field applied.

Fig. 2b shows the distributed signal S and P of section A and B after PC optimization. With the ROGUE's significant increase in back-scattering, the P-signal is increased by 45 dB compared to the bare spun fiber. According to Eq. 3, this should give a significant advantage to the ROGUE fiber in terms of noise level, which becomes very apparent when comparing the resulting magnetic field calculation on both FUT, as shown in Fig. 2c. The disadvantage of a large Δz also becomes obvious on the edge of the ROGUE, where information becomes corrupted without noise from beyond.

Result of distributed field measurement in Mag1 for a $\Delta z = 2$ cm is shown in Fig. 3a. The field between the coils is constant as should be expected in a Helmholtz configuration. The resulting PSF for a fictive highly localized field is shown in Fig. 3a. This PSF is used to define the $\Delta z$ as its FWHM. Calibration of the Verdet constant, which may vary from fiber to fiber, is done by comparing results of Eq. 2 with the calibrated measured field value. The fiber must be straight along the axial axis of the coils to ensure a measurement of the same field as that of the calibrated probes. The average Faraday rotation angle derivative in the constant region of the Helmholtz coils is shown for various expected magnetic fields in Fig. 3b. for both Mag1 and Mag2. The corresponding Verdet constant is 0.63 rad/m/T, close to previous values in the literature [9]. It is important to note that linearity is well persevered for high fields, where P and S get to similar values. Hence, in this case, we can consider that the weak linear birefringence has little impact on the quality of the measurement.

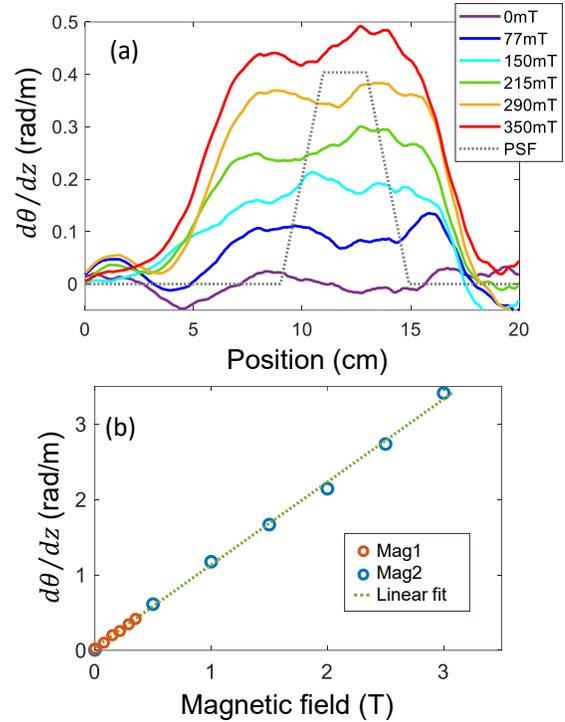

Fig. 3: (a) Measured signal with a spatial filtering of 4 cm (FWHM) for various fields in EM1. A point-spread function (PSF) of a fictive highly localized field shows the spatial filtering width. (b) Peak signal (averaged over 5 cm) in center of electromagnet for EM1 (water-cooled coil) and EM2 (superconducting magnet). The linear fit correspond to a Verdet constant of 0.63 rad/m/T.

As the noise level is highly related to the chosen Δz, an analysis of the impact of spatial filtering on the noise is shown in Fig. 4. Results of a ROGUE are compared to the unexposed spun fiber, where an improvement of one order of magnitude is shown. Noise level of 10 mT can be reached with integration length over 8 cm.

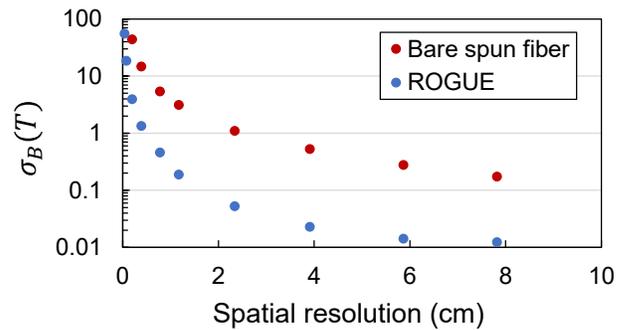

Fig. 4: RMS field noise floor vs spatial filtering length (distributed sensing spatial resolution) in regular spun fiber and in ROGUE.

Being able to measure magnetic field also allows us to measure electrical current. To do so, the optical fiber needs to be coiled around the current source cable. If we consider an integration of $N$ loops of circumference $C$, the current can be measured as:

$$I(z_e) = \cos\left(\frac{\Lambda}{C}\right)\frac{1}{N}\oint_{NC} H(z')dz'\bigg|_{z_e}, \quad (6a)$$

$$I(z) = \cos\left(\frac{\Lambda}{C}\right)\frac{\theta(z+0.5NC) - \theta(z-0.5NC)}{2VN\mu_0}, \quad (6b)$$

where $C$ is the loop circumference, $z_e = z\Lambda/C$ is the position in the current cable's axis and $\Lambda$ is the loop pitch. The cosine term in front of Eq. 6 takes into account projection of the field on the optical fiber due to a non-zero pitch and can be approximated to 1 for small pitch. This expression is valid for a constant loop pitch. At integer values of loops lengths ($z = mC$), the expression can be simplified to $I(z_e) = CH(z_e)$, where $H$ is calculated in the same way as Eq. 2 and 5. Therefore, the current noise floor can be expressed in relation to Eq. 4-5, in the approximation of small S/P ratio signal and pitch:

$$\sigma_I = \frac{C}{\mu_0}\sigma_B, \quad (7)$$

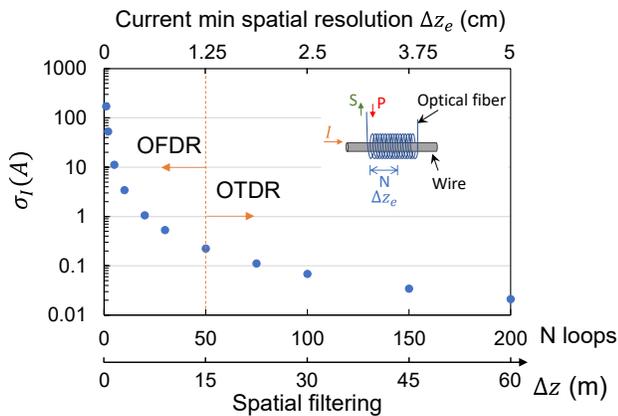

Fig. 5: Simulated current noise floor vs number of integrated loops ($\Delta z = NC$, $C$ is circumference) based on the observed experimental noise level. It is supposed that fiber is looped in circles around the current source with a diameter of 10 cm ($C = 31.4$ cm). The minimum $\Delta z_e$ is calculated for a minimum loop separation of 250 μm. 50 loops corresponds to an integration length of $\Delta z = 15$ m which is considered a practical limit for traditional OFDR, beyond which OTDR techniques should be used for long fiber length measurements.

To get an idea of how such ROGUE could potentially perform for current sensing, we extrapolated the noise modeled in Fig. 4 to large $\Delta z = NC$ and calculated equivalent current noise level in Fig. 5. Considering that this would correspond to a current distributed measurement resolution of $\Delta z_e = N\Lambda$, we associated a $N$ integrated loop number to the minimum possible resolution using a pitch of 250 μm (optical fiber diameter). As can be seen in Fig. 5, for OFDR typical sensing length (~50 m), sub-amp current could be resolved for $\Delta z = 15$ m resolution ($\Delta z_{e\ min} = 1.25$ cm). Further improvement would require using time-domain reflectometry (OTDR) instead to probe longer lengths of ROGUES.

In conclusion, we've demonstrated the use of continuous random fiber Bragg grating (ROGUE) as a magnetic distributed sensor using Faraday rotation. The higher signal to noise ratio of ROGUE yields (40 dB) a significantly improved performance in terms of measured magnetic noise floor compared to unexposed spun fiber, the standard fiber used for magnetic measurements. The higher SNR of ROGUE makes it significantly more robust to pre-FUT component losses and enables faster measurements with less gain. By extrapolating our results to longer lengths, we can predict that ROGUEs can reach high current sensitivity (~100 mA or less) in distributed measurements, enabling precise fault detection in electrical installations. Ideal ROGUEs should be written in spun fiber in order to reduce the influence of parasitic linear birefringence. This presents a challenge in the writing process due to periodic rotating stress rods, which will be tackled in future work.

**Funding.** Institutional PIED program of Polytechnique Montreal

**Acknowledgements.** Access to calibrated electromagnets was provided by Prof. David Ménard (Engineering Physics) and Prof. Frederic Sirois (Electrical Engineering) of Polytechnique Montreal.

**Disclosures.** The authors declare no conflicts of interest.